\documentstyle[12pt]{article}

\newcommand\ion[2]{#1\,\lowercase{{\sc #2}}}
\def\ref{\par\hangindent=0.375in\hangafter=1}

\begin{document}

\begin{center}
ORBITS OF HYADES MULTIPLE-LINED SPECTROSCOPIC BINARIES\\

\vspace{0.25in}
PAPER 2: THE DOUBLE-LINED SYSTEM HD 27149\\

\vspace{0.25in}
{\em By J. Tomkin\\
McDonald Observatory, Univ. of Texas at Austin}\\

\vspace{0.25in}
to appear in the February 2003 issue of {\em The Observatory}\\
\end{center}

\vspace{0.5in}
    A new determination of the orbit of the Hyades double-lined
spectroscopic binary HD 27149 is presented.  The well-defined
orbit provides the spectroscopic basis
for an extremely accurate orbital parallax
for the system --- in particular, the size of the relative orbit
($a\sin i = (a_1 + a_2)\sin i = (67.075 \pm 0.045) \times 10^6$ km)
is accurate to $\pm 0.07$\,\%.
The minimum masses for the primary and
secondary --- $m_1\sin ^3i = 1.096 \pm 0.002\,M_\odot$ and
$m_2\sin ^3i = 1.010 \pm 0.002\,M_\odot$ --- are unexpectedly
large for the spectral types thus suggesting the possibility
of eclipses.  Although the probability of eclipses is not large,
the system being composed of G3V and G6V stars in a 75-day
orbit, the possibility is of great interest.
A rediscussion of a search for eclipses made by J\o rgensen \& Olsen$^1$
in 1972 shows that central eclipses can be excluded, but that
shorter duration off-centre eclipses cannot be ruled out.
Ephemerides for possible
primary and secondary eclipses are given.

\vspace{0.25in}
\parindent 0in
{\em Introduction}
\parindent 0.3in

These papers report the determination of new orbits for
double-lined spectroscopic binaries in the Hyades cluster.
When complemented by visual orbits determined by means of
optical interferometry, which may soon start becoming
available, the results will provide accurate distances,
by means of orbital parallax, for these systems and, thus,
the Hyades cluster.  In this paper we examine HD 27149.

  Basic data for HD 27149, which has a Hyades cluster
designation vB 23, are as follows:
$\alpha$ (2000) $4^{\rm h}$ $18^{\rm m}$ $01.8^{\rm s}$,
$\delta$ (2000) $+18^\circ$ $15'$ $24''$;
spectral type G5 (from HD catalogue); $V = 7.53$, $B-V = 0.68$.
An earlier spectral type of dG3 is given by Wilson$^2$ in the
{\em General Catalogue of Stellar Radial Velocities}.  The secondary
is not much fainter than the primary and, at the low or
medium dispersions generally used for spectral typing,
the two spectra are always blended so the spectral type is
much more uncertain than that of a single star and is not
the spectral type of the primary, but is intermediate between
those of the primary and secondary.  Batten \& Wallerstein$^3$
have used the $B-V$ and $U-B$ colours and the magnitude difference
between the primary and secondary to indirectly estimate
individual spectral types of G4V and G8V.  More recently
Barrado y Navascu\'es \& Stauffer$^4$ have used the same method
to estimate individual spectral types of G3V and G6V.
As Hyades cluster members go, the system is quite bright
and may well be within the reach of the newest optical
interferometers, such as the CHARA telescope now in operation
at Mt Wilson.  In the infrared it will almost certainly
be accessible to these telescopes --- Patience {\em et al.}$^5$
use its $B$ and $B-V$ to estimate $K = 5.88$.

The cluster membership of HD 27149 is not in doubt 
--- van Bueren$^6$ identified it as a member in his early major
investigation of the cluster; more recently Griffin {\em et al.}$^7$,
in their comprehensive radial-velocity
survey of the cluster, identified it as a member in their list
of spectroscopic binaries awaiting orbit
determinations; Perryman {\em et al.}$^8$ likewise
recognise it as a member in their recent study of the cluster
based on parallaxes and proper motions measured with the
{\em Hipparcos} satellite; and Madsen {\em et al.}$^9$, who also
used the {\em Hipparcos} data but with more stringent selection
criteria for membership than Perryman {\em et al.}, include it
in their very recent list of members.
We now briefly review the history of HD 27149 as a
spectroscopic binary.

Wilson$^{10}$ discovered the spectroscopic-binary nature of HD 27149
during a survey, mostly carried out between
1942 and 1946, of the radial velocities
of 204 known and candidate Hyades cluster members, one of which
was HD 27149.  In the main table of his paper Wilson gave a mean
radial velocity for HD 27149 of +44.0 km s$^{-1}$
and classified it as a
cluster member.  The four individual velocities, given in the
footnotes to the table, are all different
and span a range of 21 km s$^{-1}$.  (The dates of the observations
are not given, but can be found in the later useful compilation
of Abt$^{11}$.)  The dispersions of Wilson's spectra --- three of
80 and one of 36\,\AA/mm --- were such that the primary and secondary
spectra of HD 27149 would have been blended so these velocities
must have been intermediate between those of the primary and
secondary and weighted towards the primary.  Wilson clearly
identified HD 27149 as a spectroscopic binary --- he stated
in the text of his paper ``[t]he probable spectroscopic binaries
number 23" and HD 27149 is one of 23 stars in the footnotes to
his main table for which he either gives individual velocities
or says two spectra are visible.
Woolley {\em et al.}$^{12}$ first detected the secondary spectrum.
Their four discovery spectra were observed on the
2.5-m telescope at Mt Wilson in October and November 1959
with a dispersion of 10\,\AA/mm.  An analysis of these spectra
by Lambert {\em et al.}$^{13}$ provided the first estimate of the
magnitude difference between the primary and secondary
--- $\Delta m = 0.69$ in $B$.  In 1973 Batten \& Wallerstein$^3$
published the first determination of the orbit.  Their analysis,
which was based on radial velocities from spectra observed with
the 5-m at Palomar, the 2.5-m at Mt Wilson and
the 1.8-m and 1.2-m telescopes at the Dominion Astrophysical
Observatory, revealed an orbital period of 75 days,
moderate eccentricity ($e = 0.23$), a mass ratio
$m_1/m_2 = 1.14$ and minimum masses ($m_1 \sin ^3i = 1.04\,M_\odot$
and $m_2 \sin ^3i = 0.91\,M_\odot$) sufficiently close to the
masses demanded by the spectral types to show that the
inclination of the plane of the orbit must be close to $90^\circ$.
Most recently, in 1982, McClure$^{14}$ reported five new observations
with the radial velocity spectrometer on the 1.2-m telescope
at the Dominion Astrophysical Observatory, which he used to
improve the Batten \& Wallerstein orbit.  And there matters have
rested with respect to our knowledge of the radial velocities
and orbit of HD 27149.

Two recent surveys of multiplicity among Hyades members
--- one by Mason {\em et al.}$^{15}$ made in 1991 with the
KPNO 4-m Mayall telescope and optical speckle imaging
and one by Patience {\em et al.}$^5$ carried out between
1993 and 1996 with the 5-m telescope at Palomar and speckle
imaging at 2.2\,$\mu$m --- both scrutinised HD 27149 and found
no sign of a visual companion.  The system, thus, consists
solely of the spectroscopic binary pair so far as we know.

This paper reports a new determination of the orbit of HD 27149
based on new high-resolution spectra from McDonald Observatory.

\vspace{0.25in}
\parindent 0in
{\em Observations and radial velocities}
\parindent 0.3in

  The observations were made on the
2.7-m and 2.1-m telescopes at McDonald Observatory between 1995 and
2002.  As shown in Table I, the first 33 observations were made
with the 2.7-m telescope and the last 17 were made with the 2.1-m.
The instrumentation and observing procedure were the same as those
used for the Hyades binary examined in the first paper in this
series$^{16}$ and have been described there.  Here it is enough to
recall that the observations on the 2.7-m telescope used
the {\em 2dcoud\'e} echelle spectrometer$^{17}$, have a
resolving power of 60\,000 and nearly complete
wavelength coverage from $\sim$\,4000 -- $\sim$\,9000\,\AA,
while those on the 2.1-m telescope used the {\em Sandiford}
Cassegrain echelle spectrograph$^{18}$, also have a resolving
power of 60\,000 and complete wavelength coverage from
$\sim$\,5600 -- $\sim$\,7000\,\AA.

  The data were processed and wavelength-calibrated in a
conventional manner with the IRAF package of programs.
The spectra are double-lined with primary and secondary
lines of similar strength and, at most orbital phases,
the secondary lines are well separated from their primary
counterparts.  Fig. 1 shows an example in which we see primary
and secondary \ion{Ni}{i} and \ion{Fe}{i} lines
near 6400\,\AA.

The procedure used to measure the radial velocities was the
same as in Paper 1 of this series and has been described there.
To summarise: the wavelengths of well-defined primary and
secondary lines were measured by fitting Gaussian profiles
with the IRAF {\em splot} routine, the wavelength differences
between the measured and rest wavelengths of the lines
provided the topocentric radial velocities, telluric O$_2$
lines were measured in the same way so as to determine the
wavelength offset between the stellar spectrum and its
associated Th--Ar comparison spectrum, the stellar
topocentric velocities were then corrected by subtracting
from them the telluric line offsets in velocity form
and, finally, the heliocentric correction led to the
heliocentric radial velocities.
The velocities are, thus, absolute velocities.

In Paper 1 a tiny additional adjustment of the stellar velocities
was made in order to force the averages for similarly measured
velocities from observations of the radial-velocity standard
$\epsilon$ Tau (G9.5\,III, itself a member of the Hyades)
on the 2.7-m and 2.1-m telescopes to be the same
as the average of all the velocities from both
telescopes; the corrections were a decrease of 70\,m s$^{-1}$
for the 2.7-m data and an increase of 140 m s$^{-1}$ for the
2.1-m.  In view of the growing realisation that K giants
are subject to intrinsic low amplitude radial-velocity variations
of $\sim$\,50 -- 400 m s$^{-1}$ over a large range of
timescales --- see, for example, Hatzes {\em et al.}$^{19}$
--- it is evident $\epsilon$ Tau might be subject
to similar radial-velocity variations
so the small difference between the velocities from
the two telescopes might be real.
There is thus no compelling reason for making an adjustment
and so it was decided not to do so
--- the measured velocities from the 2.7-m and 2.1-m
telescopes are used as they stand.  Table I gives the UT dates,
heliocentric Julian dates and heliocentric radial velocities
for the McDonald observations.  We now turn to the
determination of the primary-secondary orbit.

\vspace{0.25in}
\parindent 0in
{\em The orbit}
\parindent 0.3in

The method of differential corrections was used to determine
the primary-secondary orbit from the primary and secondary
velocities.  A necessary preliminary to the orbit calculation
was the assigment of suitable weights for the various
velocities --- namely velocities for observations with
blended primary and secondary spectra,
velocities from the 2.7-m telescope {\em versus} those from
the 2.1-m and primary {\em versus} secondary velocities.

Observations with blended primary and secondary spectra:
In a few observations the small wavelength separation between
the primary and secondary spectra meant
the primary lines and their secondary
counterparts were blended, to a greater or
lesser degree.  For these observations the
{\em deblend} option in {\em splot} had been used to fit
double Gaussian profiles to the pairs of blended primary
and secondary lines and successfully measure separate
primary and secondary velocities.  For three observations,
however, the primary-secondary wavelength separations were
so small and the blending so severe that it proved best to
give zero weight to the velocities from these observations.
In one of these cases the observation was so close to
a single-lined phase that only a single velocity representative
of the blended primary and secondary spectra could be
measured (see Table I and Fig. 2).  In the other two
observations (see Table I) the use of {\em deblend} had
provided separate primary and secondary velocities, but a
trial orbital solution showed their $O - C$ were
unusually large indicating the residual malign influence of
blending.

2.7-m {\em versus} 2.1-m telescope:  A trial orbital solution
in which the velocities from the 2.7-m and 2.1-m observations
had equal weight showed that the velocity residuals are somewhat
larger for the 2.1-m observations, which is not surprising when
one recalls that the 2.7-m observations were made at the
{\em coud\'e} focus while those on the 2.1-m were
made at the Cassegrain focus.  The criterion that the weight
multiplied by the average of the residuals squared should be
equal for the 2.7-m and 2.1-m velocities requires the weight
of the 2.1-m velocities to be $0.3\times$ the weight of the
2.7-m velocities.

Primary {\em versus} secondary velocities:  The secondary
lines are slightly weaker than the primary lines so one expects
the weights for the secondary velocities to be somewhat less
than those for the primary velocities.
In a trial solution, in which the primary and
secondary velocities had equal weight,
it was found that for the 2.7-m observations the averages of
the residuals squared for the primary and secondary velocities
demanded that the weight of the secondary velocities be
$0.5\times$ that of the primary velocities, while for the
2.1-m observations the averages of the residuals squared
for the primary and secondary velocities demanded that the
weight of the secondary velocities be $1.8\times$ that of the
primary velocities.
The cause of the unrealistic secondary
weight indicated by the 2.1-m velocity residuals
could, perhaps, be due to the small number of
2.1-m observations (15 with non-zero weight)
and the fact that a relatively large number (6) of them
happen to have been made when there was some blending
of the primary and secondary lines.  A recalculation of the
weights for primary and secondary velocities for the
2.1-m residuals using only observations in which the
velocity separation between primary and secondary is
30 km s$^{-1}$ or more gives a much more reasonable
weight of 0.95 for secondary velocities relative to
primary velocities.
It was decided to adopt the
result required by the 2.7-m observations and make the
secondary velocities half the weight of the primary velocities for both the
2.7-m and 2.1-m observations, so
the 2.7-m primary and secondary velocities have weights 1 and 0.5,
respectively, and the 2.1-m primary and secondary
velocities have weights 0.3 and 0.15, respectively.

With the weights of the various categories of velocity fixed, a new
solution of the 2.7-m and 2.1-m velocities was done.  It gave
excellent agreement of the observed and calculated velocities
--- the r.m.s. residual for the velocities of unit weight
({\em ie} 2.7-m primary velocities) is only 0.06 km s$^{-1}$.
Within reasonable limits the orbital solution is impervious to
the relative weights of the primary and secondary velocities;
for example, if instead of making the secondary velocities
half the weight of the primary velocities, we were to adopt
equal weights for the 2.7-m primary and secondary velocities
and do likewise for the 2.1-m primary and secondary velocities
the corresponding changes in the orbital elements are all much
less than the estimated errors of the orbital elements.
With the solution of the McDonald velocities accomplished,
the next step is to assemble and assess the velocities of
HD 27149 available in the literature and see if a solution
of these velocities and the McDonald velocities combined
is an improvement over the solution of the McDonald velocities
alone.

The previously published primary and secondary radial velocities
of HD 27149 are those of Woolley {\em et al.}$^{12}$
(4 observations made at Mt Wilson), those of Batten and
Wallerstein$^3$ (15 observations from Palomar or Mt Wilson, 18 from
the Dominion Astrophysical Observatory, and 1 from the Lick
Observatory) and those of McClure$^{14}$ (5 observations from the
Dominion Astrophysical Observatory).  With the exception
of McClure's radial velocities which were measured with a
radial velocity spectrometer, all of these radial velocities
were measured by means of photographic spectra.  A trial
orbital solution of these and the McDonald radial velocities
together indicated that these radial velocities, because of
their lower accuracy compared with the McDonald ones, have very
low weight --- about one thousandth for the photographic
velocities and about one hundredth for McClure's velocities.
This idea, therefore, was dropped and the
solution of the McDonald velocities alone, described above,
was adopted as the final solution.

The phases and velocity residuals for this solution are given
in Table I, the orbital elements are given in Table II and
Fig. 2 shows the observed radial velocities and calculated
radial velocity curves.  We now examine the
most interesting features of the orbit.

\vspace{0.25in}
\parindent 0in
{\em Discussion}
\parindent 0.3in

The new orbit confirms the characteristics of HD 27149 discovered
by Batten \& Wallerstein$^3$ --- a period of 75 days, moderate
eccentricity and a longitude of periastron ($\omega = 178.32 \pm 0.15$\,
degrees) which is very close to 180\,degrees so the major axis lies
almost across the line of sight with periastron passage very close
to the descending node.  The new orbit and precise orbital
parameters promise to provide, when complemented by a visual
orbit of similar quality, accurate masses for the two components
of HD 27149 and an accurate orbital parallax for the system;
the linear separation of the primary
and secondary --- $a\sin i = (a_1 + a_2)\sin i = 
(67.075 \pm 0.045) \times 10^6$ km --- is accurate to $\pm 0.07$\,\%.
This linear separation combined with a distance to HD 27149
of 47.6 pc$^5$ lead to a corresponding angular separation
of 9.4 mas for the visual orbit.

The new systemic velocity --- $\gamma = 38.461 \pm 0.014$
km s$^{-1}$ --- is little different from Batten \& Wallerstein's
result --- $38.0 \pm 0.3$ km s$^{-1}$ --- so the new orbit
does not change the radial velocity contribution to HD 27149's
kinematic situation very much.  It is of interest, however,
to compare the systemic velocity of HD 27149 with
Madsen {\em et al.'s}$^9$ recent determination of its
astrometric radial velocity.

Astrometric radial velocities, which are determined from
proper motions and trigonometric parallaxes along with the
assumption that the stars in the cluster share the same
velocity vector, are entirely independent of spectroscopy.
Madsen {\em et al.}, who identify HD 27149 by its Hipparcos
Catalogue number (20056), estimate its astrometric
radial velocity to be $38.46 \pm 0.6$ km s$^{-1}$, which
compares with the spectroscopic radial velocity of
$38.461 \pm 0.014$ km s$^{-1}$ determined here.
The excellent agreement of these two completely independent
results is most pleasing, although the sizes of the errors,
especially in the astrometric radial velocity, suggest that
the exactness of the agreement is at least partially
fortuitous.  Also it must be recognised that
the spectroscopic radial velocity includes
two effects --- the gravitational redshift and the convective
blueshift --- which do not complicate the astrometric radial
velocity.  The gravitational redshifts of the two G dwarf
components of HD 27149 must be similar to the 0.64 km s$^{-1}$
gravitational redshift of the Sun$^{20}$.  The convective
blueshift, which is an effect of stellar surface convection,
arises because the hotter rising convective cells make a larger
contribution to the total light from the stellar disk than the
cooler sinking cells.  The shift is dependent on the strength
of the absorption lines being used for radial velocity
measurement, being most marked for weak lines and almost
non-existent for very strong lines. 
From Figure 2 of Allende Prieto \& Garc\'{\i}a L\'opez$^{20}$
one finds that for the mostly medium strength
lines (intrinsic equivalent widths of between 50 and 100\,m\AA)
used to measure the radial velocities of the primary and
secondary of HD 27149 the solar convective blueshift,
which we use as an approximation to the situation in the
components of HD 27149, is $\sim 0.30 \pm 0.05$ km s$^{-1}$.
Although it might appear that the next step is to
remove the gravitational redshift and convective blueshift
from the spectroscopic radial velocity of HD 27149
so as to obtain a modified spectroscopic radial velocity
that is directly comparable with the astrometric radial
velocity, one more factor must first be reckoned with.
We recall that each individual spectroscopic radial velocity
is derived from the differences between the measured and rest
wavelengths for the set of measured stellar absorption lines
and, furthermore, the rest wavelengths
adopted for the lines are their measured wavelengths in the
Sun$^{21}$.  These solar wavelengths include the solar
gravitational redshift and convective blueshift.
The spectroscopic radial velocity of HD 27149, thus, not only
has the gravitational redshift and convective blueshift
of HD 27149 added to it but also has the solar gravitational
redshift and convective blueshift subtracted from it.
The gravitational redshifts of the primary and secondary of
HD 27149 and the Sun must be quite similar, because all three
stars lie on the same part of the main sequence, and likewise
the convective blueshifts of the three stars must also be
quite similar so for both types of shift the stellar and solar
contributions to the spectroscopic radial velocity essentially
cancel each other.  The gravitational redshift and convective
blueshift thus have little effect on the spectroscopic radial
velocity of HD 27149 and so allowance for their presence hardly
changes the excellent agreement between the spectroscopic
and astrometric radial velocities noted above.
Next we look at the minimum masses of the primary and secondary
and the possibility of eclipses.

The minimum masses of the primary and secondary are similar to
the actual masses inferred from the spectral types which suggests
the possibility of eclipses --- a circumstance already recognised
by Batten \& Wallerstein$^3$.  In fact the revised minimum masses
of the primary and secondary
--- $m_1\sin ^3i = 1.096 \pm 0.002\,M_\odot$
and $m_2\sin ^3i = 1.010 \pm 0.002\,M_\odot$ --- are slightly
larger than those determined by Batten \& Wallerstein
--- $m_1\sin ^3i = 1.04\,M_\odot$ and $m_2\sin ^3i = 0.91\,M_\odot$
--- so the case for eclipses is underlined.
We note that the minimum masses are in fact somewhat {\em larger}
than the actual masses indicated by the spectral types.
For example, with the primary and secondary spectral types of
G3 and G6, respectively, recommended by Barrado y Navascu\'es
\& Stauffer$^4$ and a standard mass {\em versus} spectral type
calibration$^{22}$ one arrives at masses of 0.98 and 0.89\,$M_\odot$
for the primary and secondary.
In view of the inherent uncertainty in spectral typing
a double-lined spectroscopic binary in which the strengths
of the primary and secondary lines are not very
different --- Batten \& Wallerstein, for example, note that an
``an (unresolved) spectrogram of dispersion 30\,\AA/mm,
obtained at Victoria, is consistent with a spectral class
between G2V and G8V" --- it is not worth making anything of
the discrepancy, but it does highlight the possibility
of eclipses.

The phases of the conjunctions, which are when eclipses will occur
if the system is eclipsing, are 0.1715 (primary in front) and
0.8369 (secondary in front).  The linear separations between
the primary and secondary at the two conjunctions are
similar --- projected onto the line of sight they are
$r \sin i = 62.94 \times 10^6$ km (primary in front)
and $61.96 \times 10^6$ km (secondary in front) --- so if there
are eclipses it is likely that there are both primary and
secondary eclipses.
If for the purposes of the present speculative discussion we
assume that the primary and secondary are both of one solar
radius, then the total duration of a central eclipse, primary
or secondary, would be $\sim 11$ hours.  The primary eclipse
would be about one magnitude deep, while the secondary eclipse
would be somewhat less deep.  None of the McDonald observations
has a phase close enough to the phases of conjunction that it
might have been made during a primary or secondary eclipse so the
spectroscopic observations are mute with respect to
the question of eclipses.  Eclipses require the
orbital inclination to be within $\sim 1^\circ .3$ of $90^\circ$
so, although their probability is small, there is still a
sporting chance of their occurrence.  We now re-examine
photometric observations of HD 27149 made in January and
February 1972 by J\o rgensen \& Olsen$^1$, who searched
for eclipses at the suggestion of Batten \& Wallerstein.

Every night from 1972 January 4 to March 1 inclusive
J\o rgensen \& Olsen observed HD 27149 once in {\em uvby} with a
simultaneous four-channel photometer on the Danish 50 cm
reflector at the European Southern Observatory.  Two nights
near conjunctions when they observed it twice and seven nights
when they did not observe it were exceptions to this routine.
There was no sign of eclipses.  But the ephemerides for
predicting conjunctions available to J\o rgensen \& Olsen
were uncertain by $\pm 1$ day so they did not know how the
conjunctions were placed with respect to their regular
once-a-night scrutiny, a circumstance which, combined with
the fact that even central eclipses would have a duration
of only about 11 hours, meant that ``any eclipse could easily
have been overlooked" --- as they pointed out.
Now, thanks to the accuracy of the new orbit, the times of
conjunction it provides can be projected back into the past
to see how the two conjunctions that occurred during
January and February 1972 were placed with respect to
J\o rgensen \& Olsen's photometric observations.
The $\pm 0.12$ day uncertainty of this backward projection
is almost entirely due to the accumulation of the period
error over the 145 orbits that separate the epoch (January 2002)
of the new orbit from 1972.  Fig.~3 compares a schematic
light curve for central eclipses with J\o rgensen \& Olsen's
photometric observations during the nights of the January and February 1972
conjunctions.  It turns out they timed their observations
quite well; central eclipses are excluded, but shorter duration
off-centre eclipses are not ruled out.
We now consider the photometry of HD 27149 provided by the {\em Hipparcos}
satellite.

This much more recent photometry consists of 41 observations, on the
{\em Hipparcos} $H_p$ magnitude system, made from January 1990 to July 1992.
Inspection of the observations, which are available at the
{\em Hipparcos} website$^{23}$, shows no sign of eclipses, but a comparison
of the calculated conjunction dates with the dates of the
{\em Hipparcos} photometry shows that all of the photometry falls
well outside the times of possible eclipses and so does not throw
any light on the eclipse question.

Ephemerides for future conjunctions are:
J.D. $2452291.8523 + 75.6587E$ (primary in front) and
     $2452266.5298 + 75.6587E$ (secondary in front).
At present and for the next four years, or so, the error in these
predicted times of conjunction is about 55 minutes.  Further into
the future the error will slowly increase as the accumulation
of the period error makes its presence felt.  Table III gives
the Julian and UT dates provided by these ephemerides for
the last three months of 2002 and all of 2003.
Photometrists who find themselves on the night side of the Earth
at these times are encouraged to turn their telescopes towards
HD 27149 and see if it eclipses.  The probability is small, but
the prospective reward is great.

In conclusion we summarise our picture of HD 27149 as follows:
The new spectroscopic orbit provides the basis
for an accurate orbital parallax, the expected angular separation
is $\sim 9$ mas, the minimum masses must be very close to the actual
masses and, although the photometric observations rule out central eclipses,
the door to off-centre eclipses is still open.

\vspace{0.25in}
\parindent 0in
{\em Acknowledgements}
\parindent 0.3in

  Observations of HD 27149 by David Lambert are gratefully acknowledged
and thanks go to Frank Fekel for providing the
orbit solution programs.  Roger Griffin pointed me towards the
{\em Hipparcos} photometry.  David Doss' and Jerry Martin's
patient guidance was essential for the successful operation of
the McDonald telescopes and their instrumentation.  This work
was supported, in part, by a grant (AST9618414) from
the National Science Foundation of the U.S.A. and by
the Cox Endowment of the Astronomy Department, University
of Texas.

\newpage
\begin{center}
{\em References}
\end{center}
\parindent=0in

\ref (1) B. G. J\o rgensen \& E. H. Olsen, {\em IBVS,} No. 652, 1972.

\ref (2) R. E. Wilson, {\em General Catalogue of Stellar Radial Velocities}
    (Carnegie Institution of Washington, Washington D. C.), 1953.

\ref (3) A. H. Batten \& G. Wallerstein, {\em PDAO,} {\bf 14}, 135, 1973.

\ref (4) D. Barrado y Navascu\'es \& J. R. Stauffer, {\em A\&A},
    {\bf 310}, 879, 1996.

\ref (5) J. Patience {\em et al., AJ,} {\bf 115}, 1972, 1998.

\ref (6) H. G. van Bueren, {\em Bull. Astron. Inst. Netherlands},
    {\bf 432}, 385, 1952.

\ref (7) R. F. Griffin {\em et al.}, {\em AJ}, {\bf 96}, 172, 1988.

\ref (8) M. A. C. Perryman {\em et al., A\&A,} {\bf 331}, 81, 1998.

\ref (9) S. Madsen, D. Dravins \& L. Lindegren, {\em A\&A}, {\bf 381},
    446, 2002.

\ref (10) R. E. Wilson, {\em ApJ}, {\bf 107}, 119, 1948.

\ref (11) H. A. Abt, {\em ApJS}, {\bf 19}, 387, 1970.

\ref (12) R. v.d. R. Woolley, D. H. P. Jones \& L. M. Mather,
    {\em ROB}, No. 23, 1960.

\ref (13) D. L. Lambert, A. R. D. Norman \& D. H. P. Jones,
    {\em The Observatory}, {\bf 81}, 145, 1961.

\ref (14) R. D. McClure, {\em ApJ}, {\bf 254}, 606, 1982.

\ref (15) B. D. Mason {\em et al.}, {\em AJ}, {\bf 105}, 220, 1993.

\ref (16) J. Tomkin \& R. F. Griffin, {\em The Observatory}, {\bf 122},
    1, 2002.

\ref (17) R. G. Tull {\em et al., PASP,} {\bf 107}, 251, 1995.

\ref (18) J. K. McCarthy {\em et al., PASP,} {\bf 105}, 881, 1993.

\ref (19) A. P. Hatzes, A. Kanaan \& D. Mkrtichian, in J. B. Hearnshaw
    \& C. D. Scarfe (eds.), {\em Precise Stellar Radial Velocities}
    (ASP Conference Series, Vol. 185), 1999, p. 166.

\ref (20) C. Allende Prieto \& R. J. Garc\'{\i}a L\'opez, {\em A\&ASS},
    {\bf 129}, 41, 1998.

\ref (21) C. E. Moore, M. G. J. Minnaert \& J. Houtgast,
    {\em The Solar Spectrum 2935\,\AA\ to 8770\,\AA},
    NBS monograph 61 (U.S. Government Printing Office, Washington D.C.),
    1966.

\ref (22) J. S. Drilling \& A. U. Landolt, in A. N. Cox (ed.)
    {\em Allen's Astrophysical Quantities} (Springer-Verlag,
    New York), 2000, p. 381.

\ref (23) {\em The Hipparcos and Tycho Catalogues},
    http://astro.estec.esa.nl/Hipparcos/hipparcos.html.

\newpage
Figure captions
\parindent 0.3in

Fig.~1:  The spectrum of HD 27149 at 6380\,\AA\ showing
a \ion{Ni}{i} line (rest wavelength 6378.3\,\AA) and a \ion{Fe}{i} line
(rest wavelength 6380.8\,\AA) in the primary and secondary.

Fig.~2:  The radial velocities for the primary and secondary of
HD 27149 and the calculated radial velocity curves.  The filled
circle near phase 0.8 is the velocity for the blended primary and
secondary spectra in an observation at a
single-lined phase; the velocity has zero weight in the solution
for the orbits.

Fig.~3:  The J\o rgensen \& Olsen photometry of HD 27149 at the
January 1972 conjunction (possible secondary eclipse) and the February
1972 conjunction (possible primary eclipse) compared with a schematic
light curve for central eclipses.  Times of mid-eclipse are
estimated by projecting the times of conjunction for the new
orbit back into the past; the dotted lines show how much earlier
or later these eclipses might have been as a result of the
0.12 day uncertainty in the projection.  The two observations
close together on the night of 1972 February 24 exclude
a central eclipse.  If the conjunctions were earlier, within the limit
set by the uncertainty in the projection, these two points would
exclude central eclipses even more firmly, while,
if the conjunctions were later, then these
two points in combination with that for the night of
1972 January 5 would also exclude central eclipses.
The J\o rgensen \& Olsen photometry, thus, rules out
central eclipses.  Shorter duration off-centre eclipses,
however, may still have managed to fit themselves into the gaps
between the photometric observations.

\clearpage
\pagestyle{empty}
%
%
\begin{table}  
\begin{center}  
T{\sc able} I\\
{\em Radial velocities of HD 27149}\\
\begin{tabular}{lrrrrrrr} 
\multicolumn{1}{c}{\em Date (UT)}  &  $Tel$  &  \multicolumn{1}{c}{$HJD$}  &  {\em Phase}
  &  \multicolumn{2}{c}{\em Velocity}  &  \multicolumn{2}{c}{$(O - C)$}\\
  &  &  ${\em -2400000}$  &  &  \multicolumn{1}{c}{$Pri$}
  &  \multicolumn{1}{c}{$Sec$}  &  \multicolumn{1}{c}{$Pri$}
  &  \multicolumn{1}{c}{$Sec$}\\
  &  &  &  &  {\em km s}$^{\em -1}$  &  {\em km s}$^{\em -1}$
  &  {\em km s}$^{\em -1}$  &  {\em km s}$^{\em -1}$\\
\\
1995 Aug 09  & 2.7 & 49938.932  &  0.0724  &  $ 6.52$   & 73.16   &  $ 0.03$  &   $ 0.02$\\
1995 Sep 30  & 2.7 & 49990.854  &  0.7587  &  $45.02$   & 31.46   &  $ 0.01$  &   $ 0.11$\\
1995 Oct 01* & 2.7 & 49991.847  &  0.7718  &  $43.16$   & 33.84   &  $ 0.26$  &   $ 0.20$\\
1995 Oct 12  & 2.7 & 50002.931  &  0.9183  &  $ 9.93$   & 69.40   &  $-0.04$  &   $ 0.04$\\
1995 Oct 13  & 2.7 & 50003.942  &  0.9317  &  $ 6.83$   & 72.59   &  $-0.06$  &   $-0.12$\\
1995 Oct 14  & 2.7 & 50004.886  &  0.9441  &  $ 4.36$   & 75.45   &  $ 0.09$  &   $-0.10$\\
1995 Oct 15  & 2.7 & 50005.914  &  0.9577  &  $ 1.71$   & 78.24   &  $-0.11$  &   $ 0.03$\\
1995 Dec 02  & 2.7 & 50053.778  &  0.5904  &  $60.52$   & 14.60   &  $ 0.04$  &   $ 0.02$\\
1995 Dec 03  & 2.7 & 50054.749  &  0.6032  &  $59.97$   & 15.20   &  $ 0.05$  &   $ 0.02$\\
1995 Dec 03  & 2.7 & 50054.896  &  0.6051  &  $59.82$   & 15.36   &  $-0.01$  &   $ 0.08$\\
1995 Dec 04  & 2.7 & 50055.752  &  0.6164  &  $59.26$   & 15.88   &  $ 0.00$  &   $-0.03$\\
1995 Dec 04  & 2.7 & 50055.918  &  0.6186  &  $59.15$   & 16.10   &  $ 0.01$  &   $ 0.06$\\
1996 Jan 04  & 2.7 & 50086.650  &  0.0248  &  $-1.18$   & 81.21   &  $-0.13$  &   $-0.10$\\
1996 Jan 05  & 2.7 & 50087.698  &  0.0387  &  $ 0.45$   & 79.58   &  $-0.03$  &   $-0.08$\\
1996 Jan 06  & 2.7 & 50088.586  &  0.0504  &  $ 2.23$   & 77.70   &  $-0.02$  &   $-0.04$\\
1996 Jan 06  & 2.7 & 50088.836  &  0.0537  &  $ 2.71$   & 77.21   &  $-0.10$  &   $ 0.08$\\
1996 Feb 06  & 2.7 & 50119.610  &  0.4605  &  $61.54$   & 13.41   &  $-0.04$  &   $ 0.02$\\
1996 Feb 07  & 2.7 & 50120.582  &  0.4733  &  $61.74$   & 13.09   &  $-0.08$  &   $-0.04$\\
1996 Feb 08  & 2.7 & 50121.579  &  0.4865  &  $61.94$   & 12.76   &  $-0.05$  &   $-0.18$\\
1996 Feb 09  & 2.7 & 50122.602  &  0.5000  &  $62.02$   & 12.83   &  $-0.06$  &   $-0.01$\\
1997 Jan 14  & 2.7 & 50462.743  &  0.9958  &  $-1.83$   & 82.36   &  $ 0.09$  &   $ 0.10$\\
1997 Jan 19  & 2.7 & 50467.693  &  0.0612  &  $ 4.21$   & 75.58   &  $ 0.02$  &   $-0.05$\\
1997 Aug 29  & 2.7 & 50689.964  &  0.9990  &  $-1.96$   & 82.40   &  $ 0.03$  &   $ 0.07$\\
\end{tabular}
\end{center}
\end{table}
\clearpage
\begin{table}
\begin{center}
T{\sc able} I (continued)\\
{\em Radial velocities of HD 27149}\\
\begin{tabular}{lrrrrrrr} 
\multicolumn{1}{c}{\em Date (UT)}  &  $Tel$  &  \multicolumn{1}{c}{$HJD$}  &  {\em Phase}
  &  \multicolumn{2}{c}{\em Velocity}  &  \multicolumn{2}{c}{$(O - C)$}\\
  &  &  ${\em -2400000}$  &  &  \multicolumn{1}{c}{$Pri$}
  &  \multicolumn{1}{c}{$Sec$}  &  \multicolumn{1}{c}{$Pri$}
  &  \multicolumn{1}{c}{$Sec$}\\
  &  &  &  &  {\em km s}$^{\em -1}$  &  {\em km s}$^{\em -1}$
  &  {\em km s}$^{\em -1}$  &  {\em km s}$^{\em -1}$\\
\\
1997 Aug 30  & 2.7 & 50690.960  &  0.0122  &  $-1.75$   & 82.19   &  $ 0.08$  &   $ 0.02$\\
1997 Aug 31  & 2.7 & 50691.948  &  0.0252  &  $-0.99$   & 81.34   &  $ 0.02$  &   $ 0.06$\\
1997 Sep 01  & 2.7 & 50692.936  &  0.0383  &  $ 0.46$   & 79.85   &  $ 0.04$  &   $ 0.13$\\
1997 Sep 02  & 2.7 & 50693.946  &  0.0516  &  $ 2.52$   & 77.60   &  $ 0.07$  &   $ 0.08$\\
1997 Dec 10  & 2.7 & 50792.874  &  0.3592  &  $56.81$   & 18.73   &  $ 0.02$  &   $ 0.15$\\
1997 Dec 13  & 2.7 & 50795.702  &  0.3966  &  $59.24$   & 16.06   &  $ 0.06$  &   $ 0.07$\\
1997 Dec 14  & 2.7 & 50796.741  &  0.4103  &  $59.84$   & 15.28   &  $-0.02$  &   $ 0.03$\\
1997 Dec 15  & 2.7 & 50797.749  &  0.4236  &  $60.50$   & 14.75   &  $ 0.06$  &   $ 0.13$\\
1998 Oct 01  & 2.7 & 51087.954  &  0.2593  &  $46.09$   & 30.28   &  $ 0.03$  &   $ 0.06$\\
1998 Oct 05  & 2.7 & 51091.928  &  0.3118  &  $52.66$   & 23.21   &  $ 0.09$  &   $ 0.05$\\
2000 Feb 03  & 2.1 & 51577.665  &  0.7320  &  $48.72$   & 27.08   &  $-0.12$  &   $-0.12$\\
2000 Oct 26  & 2.1 & 51843.796  &  0.2495  &  $44.74$   & 31.80   &  $ 0.14$  &   $-0.01$\\
2000 Nov 11  & 2.1 & 51859.848  &  0.4616  &  $61.72$   & 13.50   &  $ 0.12$  &   $ 0.14$\\
2000 Nov 12  & 2.1 & 51860.742  &  0.4735  &  $61.64$   & 12.89   &  $-0.18$  &   $-0.23$\\
2000 Nov 13  & 2.1 & 51861.746  &  0.4867  &  $62.04$   & 12.88   &  $ 0.05$  &   $-0.06$\\
2000 Dec 07* & 2.1 & 51885.704  &  0.8034  &  \multicolumn{2}{c}{38.44}  &  $ 1.28$  &   $-1.44$\\
2000 Dec 08* & 2.1 & 51886.652  &  0.8159  &  $34.26$   & 42.43   &  $-0.35$  &   $-0.20$\\
2000 Dec 09  & 2.1 & 51887.678  &  0.8295  &  $31.55$   & 45.93   &  $-0.15$  &   $ 0.13$\\
2001 Jan 11  & 2.1 & 51920.669  &  0.2655  &  $46.69$   & 29.16   &  $-0.25$  &   $-0.11$\\
2001 Jan 12  & 2.1 & 51921.622  &  0.2781  &  $48.33$   & 27.33   &  $-0.30$  &   $-0.10$\\
2001 Oct 30  & 2.1 & 52212.918  &  0.1282  &  $19.62$   & 58.39   &  $-0.23$  &   $-0.26$\\
2001 Oct 31  & 2.1 & 52213.947  &  0.1418  &  $23.08$   & 55.03   &  $-0.08$  &   $-0.02$\\
2001 Nov 01  & 2.1 & 52214.886  &  0.1543  &  $26.13$   & 51.82   &  $ 0.02$  &   $-0.04$\\
\end{tabular}
\end{center}
\end{table}
\clearpage
\begin{table}
\begin{center}
T{\sc able} I (conclusion)\\
{\em Radial velocities of HD 27149}\\
\begin{tabular}{lrrrrrrr} 
\multicolumn{1}{c}{\em Date (UT)}  &  $Tel$  &  \multicolumn{1}{c}{$HJD$}  &  {\em Phase}
  &  \multicolumn{2}{c}{\em Velocity}  &  \multicolumn{2}{c}{$(O - C)$}\\
  &  &  ${\em -2400000}$  &  &  \multicolumn{1}{c}{$Pri$}
  &  \multicolumn{1}{c}{$Sec$}  &  \multicolumn{1}{c}{$Pri$}
  &  \multicolumn{1}{c}{$Sec$}\\
  &  &  &  &  {\em km s}$^{\em -1}$  &  {\em km s}$^{\em -1}$
  &  {\em km s}$^{\em -1}$  &  {\em km s}$^{\em -1}$\\
\\
2001 Nov 25  & 2.1 & 52238.806  &  0.4704  &  $61.73$   & 13.11   &  $-0.04$  &   $-0.07$\\
2001 Dec 31  & 2.1 & 52274.679  &  0.9446  &  $ 4.14$   & 75.65   &  $-0.04$  &   $ 0.01$\\
2002 Feb 01  & 2.1 & 52306.648  &  0.3671  &  $57.34$   & 17.97   &  $-0.02$  &   $ 0.01$\\
2002 Feb 02  & 2.1 & 52307.678  &  0.3807  &  $58.18$   & 17.00   &  $-0.08$  &   $ 0.01$\\
\\
\end{tabular}
\end{center}
The 2.7-m primary and secondary velocities have weights 1 and 0.5,
respectively; the 2.1-m primary and secondary velocities
have weights 0.3 and 0.15, respectively.
The velocities for the three asterisked observations
have zero weight; one of these observations --- the one with only
one velocity, representing the blended primary and secondary
spectra --- is single-lined and the other two are nearly so.
\end{table}

\clearpage
%
%
\begin{table}
\begin{center}
T{\sc able} II\\
{\em Orbital elements of HD 27149}\\
\begin{tabular}{llll}
\\
$P$ (days)  &  $ = 75.6587 \pm 0.0008$
  &  ~~~~~~~~~~~~~$T$ (HJD)  &  $ = 2452278.874 \pm 0.036$\\
$\gamma$ (km s$^{-1}$)  &  $ = 38.461 \pm 0.014$
  &  ~~~~~~~~~~~~~$q$ ($=m_1/m_2$)  &  $ = 1.085 \pm 0.001$\\
$K_1$ (km s$^{-1}$)  &  $ = 32.054 \pm 0.019$
  &  ~~~~~~~~~~~~~$m_1\sin ^3i$ ($M_\odot$)  &  $ = 1.096 \pm 0.002$\\
$K_2$ (km s$^{-1}$)  &  $ = 34.767 \pm 0.040$
  &  ~~~~~~~~~~~~~$m_2\sin ^3i$ ($M_\odot$)  &  $ = 1.010 \pm 0.002$\\
$e$  &  $ = 0.2628 \pm 0.0006$
  &  ~~~~~~~~~~~~~$a_1\sin i$ ($10^6$\,km)  &  $ = 32.176 \pm 0.020$\\
$\omega$ (degrees)  &  $ = 178.32 \pm 0.15$
  &  ~~~~~~~~~~~~~$a_2\sin i$ ($10^6$\,km)  &  $ = 34.899 \pm 0.040$\\
\\
\end{tabular}
R.m.s. residual (unit weight) = 0.06 km s$^{-1}$
\end{center}
\end{table}

\clearpage
%
%

\begin{table}
\begin{center}
T{\sc able} III\\
{\em Conjunctions of HD 27149, October 2002 to end of 2003}\\
\begin{tabular}{rrc}
\\
\multicolumn{1}{c}{JD}  &  \multicolumn{1}{c}{UT date}  &  Pri/sec\\
    &           &  in front\\
2452569.16  &  2002 Oct 21.66  &  sec\\
2452594.49  &       Nov 15.99  &  pri\\
2452644.82  &  2003 Jan 05.32  &  sec\\
2452670.15  &       Jan 30.65  &  pri\\
2452720.48  &       Mar 21.98  &  sec\\
2452745.80  &       Apr 16.30  &  pri\\
2452796.14  &       Jun 05.64  &  sec\\
2452821.46  &       Jun 30.96  &  pri\\
2452871.80  &       Aug 20.30  &  sec\\
2452897.12  &       Sep 14.62  &  pri\\
2452947.46  &       Nov 03.96  &  sec\\
2452972.78  &       Nov 29.28  &  pri\\
\\
\end{tabular}
\end{center}
Dates are heliocentric --- the difference between heliocentric
and geocentric dates is insignificant compared to the 55 minute
uncertainty of the predictions.
\end{table}

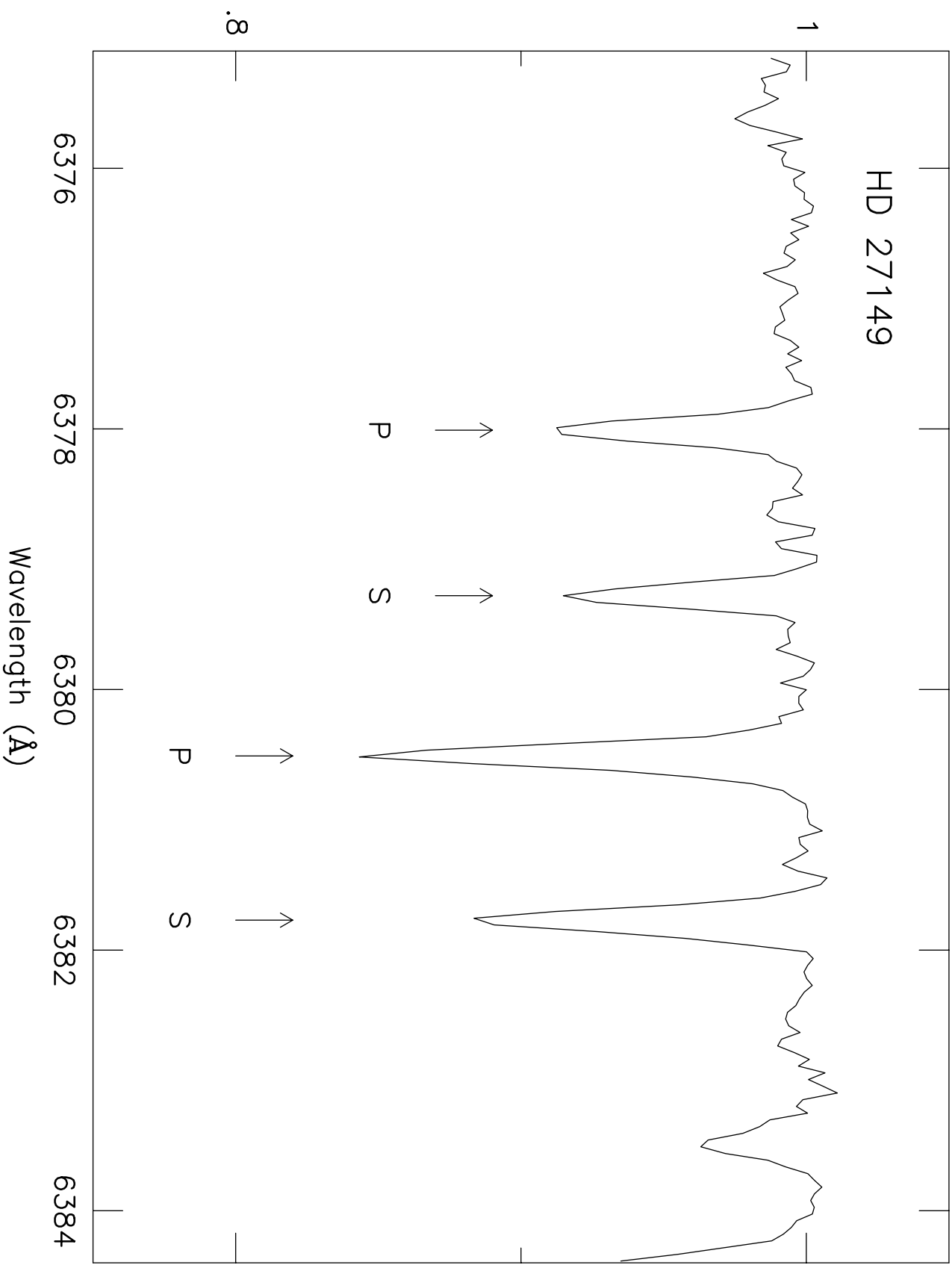
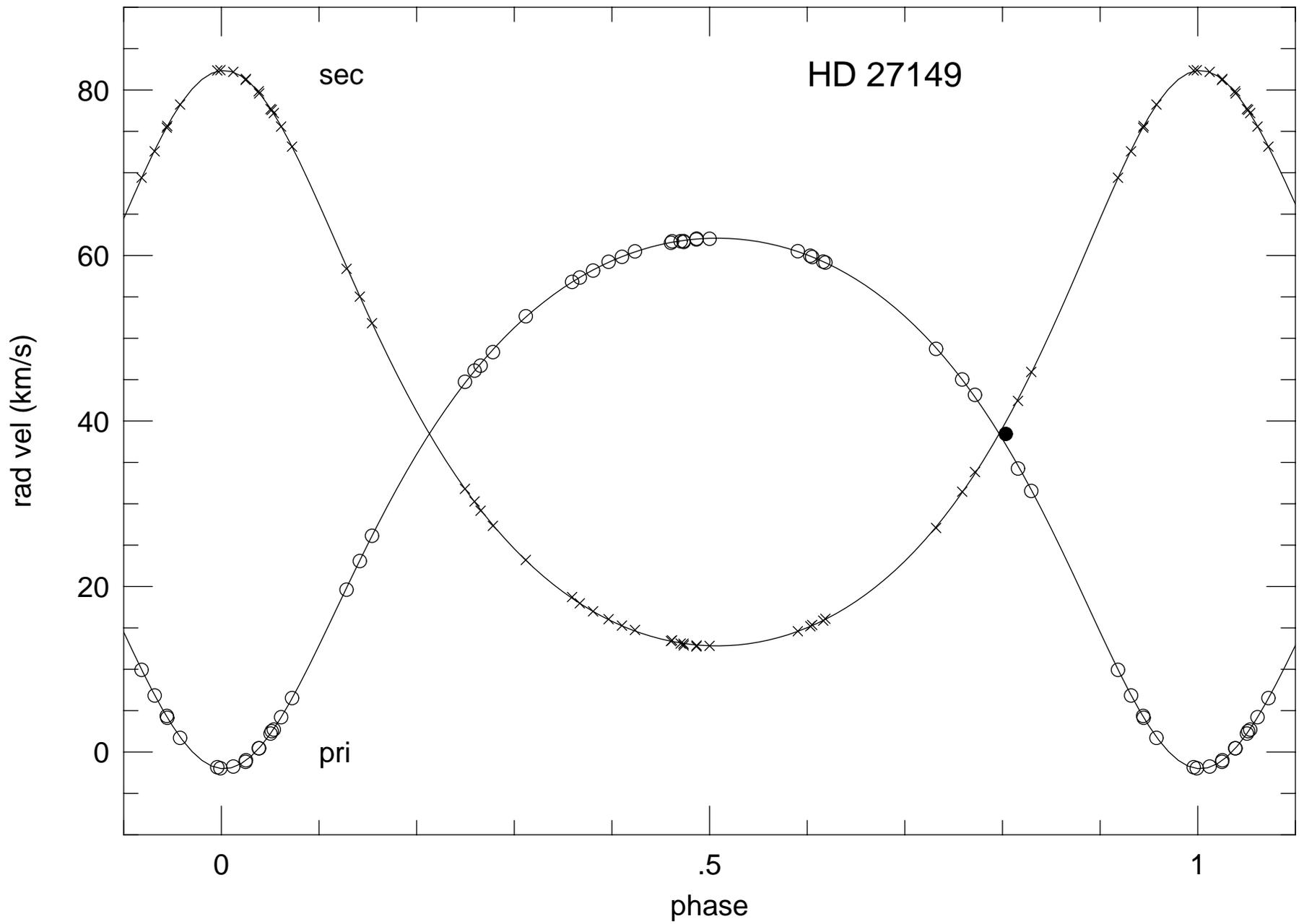
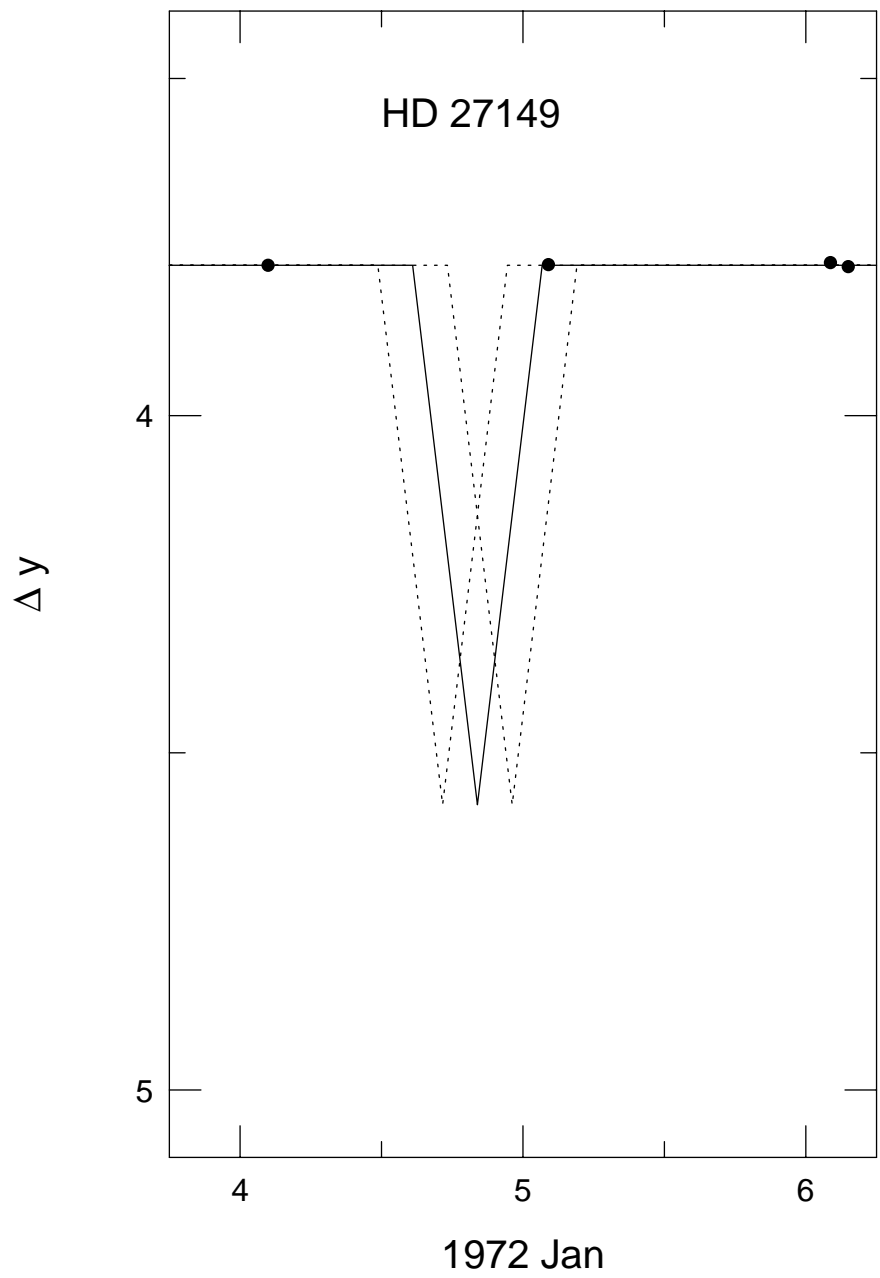
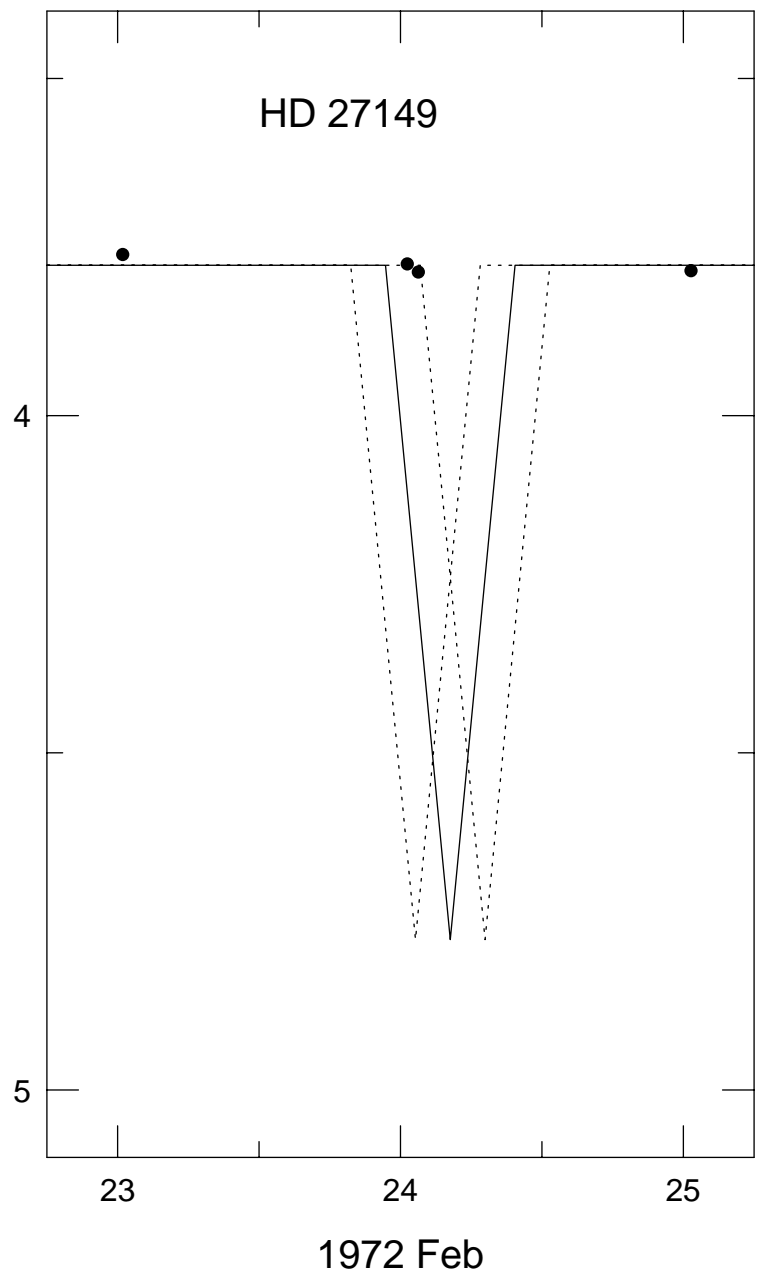
\end{document}